\documentclass[a4paper,11pt]{article}
\pdfoutput=1 % if your are submitting a pdflatex (i.e. if you have
\usepackage{jheppub} % for details on the use of the package, please
\usepackage[utf8]{inputenc}                    
\usepackage[T1]{fontenc} % if needed
\usepackage{amsmath}
\newcommand{\gj}{$\gamma + 3 \mathrm{jets}$}
\newcommand{\seff}{$\sigma_{\textit{eff}}$}
\newcommand{\seffcdf}{$\sigma_{\textit{eff,CDF}}$}
\newcommand{\seffcdfnew}{$\bar\sigma_{\textit{eff,CDF}}$}

\newcommand\Pythia{\textsf{Pythia~6}}
\newcommand\PythiaL{\textsf{Pythia~6.4.26}}
\newcommand\Herwig{\textsf{Herwig++}}
\newcommand\HerwigL{\textsf{Herwig++~2.5.2}}
\newcommand\fHerwig{\textsf{fHerwig}}
\newcommand\fHerwigL{\textsf{fHerwig~6.510}}
\newcommand\F{f}
\newcommand\CDFJetClu{\textsf{CDFJetClu}}
\newcommand\AntiK{\textsf{Anti-k$_T$}}
\newcommand\PxCone{\textsf{PxCone}}
\newcommand\MRST{\textsf{MRST98}}
\newcommand\CTEQ{\textsf{CTEQ6L1}}
\newcommand\CTEQL{\textsf{CTEQ5L}}
\newcommand\MRSTLO{\textsf{MRST~LO**}}
\newcommand\R{\mathcal{R}}

\title{\boldmath Extracting \seff{} from the CDF \gj{} measurement}

\author[a]{M. B\"ahr,}
\author[b]{M. Myska,}
\author[c]{M. H. Seymour}
\author[c]{A. Si\'odmok}

\affiliation[a]{Blue Yonder GmbH \& Co. KG,
                \\Karlsruher Strasse 88, 76139 Karlsruhe, Germany}
\affiliation[b]{FNSPE, Czech Technical University in Prague,
		\\Brehova 7, 115 19, Prague, Czech Republic}
\affiliation[c]{Consortium for Fundamental Physics,
		School of Physics and Astronomy, 
		\\ The University of Manchester, Manchester, M13 9PL, U.K.}

\emailAdd{miroslav.myska@fjfi.cvut.cz}
\emailAdd{michael.seymour@manchester.ac.uk}
\emailAdd{andrzej.siodmok@manchester.ac.uk}

\abstract{
In their 1997 paper, CDF measured \seff{}, the normalization factor that relates 
the cross section for double parton scattering to the product of the inclusive cross
sections of the two individual scatters, in a model in which they are assumed to 
be independent. In his 2007 paper, Treleani pointed out that CDF used a non-standard 
definition,  in which the double parton scattering cross section corresponds to 
\emph{exactly\/} two scatters, rather than the more conventional one in which it is 
the \emph{inclusive\/} two-scatter cross section.  He also estimated the correction 
from one definition to the other, to give a corrected value of \seff{}. Treleani's 
form would be correct under the assumption that CDF were able to uniquely identify 
and count the number of scatters in an event, which is certainly not the case. In 
this publication we consider CDF's event definition in more detail to provide an 
improved correction. 
}
\preprint{{\flushright MAN/HEP/2013/02 \\ MCnet-13-01\\ }}
\begin{document}
\maketitle
\flushbottom
\section{Introduction}
\label{sec:intro}
Due to the composite nature of hadrons, it is possible to have multiple parton
scatterings, i.e.\ events in which two or more distinct parton 
interactions occur simultaneously in a single hadron-hadron collision. At fixed 
final state invariant masses, such cross sections tend to increase with 
collision energy because partons with successively lower momentum fraction $x$, 
hence rapidly increasing fluxes, are probed. Therefore, the question of 
multiple parton interactions (MPI) in a single hadronic collision has rapidly moved 
from a theoretical curiosity, when double parton interactions was for the 
first time observed by  AFS experiment at $\sqrt{s} = 63$~GeV~\cite{Akesson:1986iv}
%in four-jet events, 
to a critical issue at the LHC. Multiple parton interactions at the LHC 
give rise to different effects, among others a substantial increase of the
unavoidable background to most observables used for the search of new physics 
\cite{Bandurin:2010gn,Hussein:2006xr,PhysRevD.61.077502,Maina:2010vh,Maina:2009sj,Godbole:1989ti}.
For this reason,
multiple parton scattering has taken on considerable importance in recent years, 
since a variety of new and improved Monte Carlo models of underlying event
physics rely on it \cite{Sjostrand:1987su, Engel:1994vs,
  Butterworth:1996zw, Borozan:2002fk, Sjostrand:2004pf,
  Sjostrand:2004ef, Hoche:2007hg, Bahr:2008dy, Gieseke:2012ft, Bartalini:2011jp}.  
Unfortunately the process cannot be estimated in a straightforward
way, due to the lack of knowledge of the non-perturbative physics, 
therefore these models rely on the experimental input.
Especially, the experimental value of \seff{}, the normalization factor that relates 
the cross section for double parton scattering to the product of the inclusive cross 
sections of the two individual scatters, has the potential to act as a strong constraint 
on models of multiple parton scattering and, in particular, their models of the 
transverse-space distribution of partons in hadrons. The value of \seff{} is also  
crucial for calculations of double pair production based on the framework of 
the $k_t$-factorization approach, see for example~\cite{Baranov:2012re}. 

CDF's measurement of the double parton scattering cross section~\cite{Abe:1997xk}, 
still one of the best available, therefore has considerable significance not only for 
MPI models but also as a \textit{standard candle} for 
new~\cite{Aaij:2012dz,Aaij201252,LHCb-CONF-2011-009}
and planned measurements~\cite{Dobson:1404953}. 
CDF's measurement was better than any that came before it, because it
avoided almost all reliance on a Monte Carlo description of their final
states and on theoretically-calculated cross sections.  Instead it made
an ingenious direct extraction of \seff{} by defining an event
selection sensitive to double parton scattering and comparing the rate
of these events from beam crossings with a single vertex (assumed to be
double parton scattering within one proton--antiproton collision) and
with two vertices (assumed to be single-parton scatterings within two
independent proton--antiproton collisions).  The main assumption that
their extraction relies on is that the final state of two scatters in
the same proton--antiproton collision is identical to that of two
scatters in different proton--antiproton collisions. CDF \emph{defined\/} 
the cross section normalization factor \seff{} through the equation
\begin{equation}
    \sigma_{ab;2} = \frac{\sigma_a\,\sigma_b}{\mbox{\seffcdf}},
      \label{ab2}
\end{equation}
where $\sigma_{ab;2}$ is the cross section for a colliding
proton--antiproton pair to have \emph{exactly\/} two scatters of types $a$ and~$b$. We choose the 
notation for exclusive cross sections, where the process is described by the subscript 
containing the type of individual parton processes and the number of scatters after 
a semi-colon. Here, we assume that the two processes $a$ and~$b$ are different,
i.e.~distinguishable\footnote{CDF use an experimental method to verify that this is 
the case with their event selection, even though there could be some overlap in principle.}. 
$\sigma_{a,b}$ are their
\emph{inclusive\/} cross sections.  To distinguish this from the
definition more commonly used in theoretical studies, we write it as
\seffcdf{} from here on. CDF's final value was
\begin{equation}
  \mbox{\seffcdf} = (14.5\pm1.7^{+1.7}_{-2.3})~\mbox{mb}.
\end{equation}
Since they used the definition of (\ref{ab2}), which refers to exactly
two scatters, they made a correction to account for the fact that a
fraction of their events (which they estimated to be 17$^{+4}_{-8}\%$) came from
triple-parton scattering events. In their analysis CDF used this estimation to re-scale 
the number of the accepted events by the factor of 0.83$^{+0.08}_{-0.04}$. The ratio
of these two numbers is equal to the ratio of the exclusive triple and double parton scattering 
cross sections producing the same final state:
\begin{equation}
 \frac{\sigma_{ab;3}}{\sigma_{ab;2}} \approx \frac{17}{83}.
 \label{eq:s3overs2}
\end{equation} 

From the theoretical point of view, it is more convenient to define
\seff{} through an analogous formula, but for the
\emph{inclusive\/} double-scattering cross section,
\begin{equation}
  \label{abincl}
  \sigma_{ab} = \frac{\sigma_a\,\sigma_b}{\mbox{\seff}},
\end{equation}
since, as shown in Section~\ref{PartonLevel}, with this definition in the assumption 
of independence of individual scatters, \seff{} depends only on properties of the
colliding hadrons and not on the scattering types or cross sections.

The non-standard definition used by CDF has been pointed out for the first time by
Treleani in his publication~\cite{Treleani:2007gi}.
Based on the theoretically-pure (parton-level) situation, assuming 
that one could measure, for a given event, whether it came from two scatters 
of definite types $a$ and $b$ or three, one of type $a$ and two of type $b$, 
Treleani calculated the correction from \seffcdf{} to \seff{}.
Using the formula in his paper, one obtains the value\footnote{Which in~\cite{Treleani:2007gi} 
is written as ``$\mbox{\seff} \approx 11\mbox{mb}$''.}
\begin{equation}
  \mbox{\seff}=10.3~\mbox{mb}.
\end{equation}

The purpose of this publication is to point out that this extraction is
over-simplified~-- the indirectness of CDF's measurement means that they
are far from being in this pure (parton-level) situation.

The paper is organised as follows. We begin by recapping the salient points 
of CDF's measurement and event selection in Sect.~\ref{CDF}. In Sect.~\ref{PartonLevel} 
we consider how to calculate the two- and three-scatter cross sections with this event
selection.  We show, in Sect.~\ref{HadronLevel}, that further input is needed to relate these to
\seff{}.  In Sect.~\ref{CorrectionFactor} we make Monte Carlo estimates
of this input and its uncertainty. One important point to note already
is that we only need ratios of closely-related cross sections, so we
hope not to be too sensitive to details of the Monte Carlo event
generation so that the uncertainty is fairly small. Finally, in
Sect.~\ref{Results} we put this input together with CDF's measurement to
give a final value for \seff{}.
In appendices we give more details backing up our Monte Carlo evaluation
of the correction factor and its uncertainty.

\section{CDF's experimental measurement}

The intricacies of the clever extraction of \seff{} using the single- and 
double-vertex data will not need to concern us here. What will be important 
is the definition of the final state containing a direct photon 
and exactly three jets. The photon was measured within pseudo-rapidity 
acceptance $|\eta|<0.9$ and was required to have $E_{T\gamma}>16$~GeV. Jets 
were reconstructed from all objects in the calorimeter region, $|\eta|<4.2$, 
using a cone algorithm \CDFJetClu{} with cone radius $0.7$~\cite{Abe:1991ui}. 
Jets were ordered according to their $E_T$ so that $E_{T1}>E_{T2}>E_{T3}$ and 
were required to be separated in $\eta-\phi$ plane from each other by 0.7 
and from the photon by $0.8$. Events were accepted if all three jets had $E_T>5$~GeV 
and simultaneously the second and the third hardest jets had $E_T<7$~GeV.

In what we will call the \textit{pure parton picture}, one would therefore have one of
the jets also above $16$~GeV, equal and opposite to the photon in
transverse momentum both coming from one scatter, and the other two
jets equal and opposite to each other, from the other scatter.
However, with $5$~GeV jets, one is far from this theoretically pure
situation.  CDF spent some time investigating the properties of such
low $E_T$ jets and concluded that they are sufficiently correlated
with the underlying parton dynamics to enable their measurement but
that there was a great deal of smearing and creation of jets `from
nothing'. We certainly cannot therefore rely on this simple parton
picture. In fact, CDF estimated that $75$\% of their event sample came
from events in which the photon and two of the jets came from one
scattering and one jet from the other. They stressed throughout that 
the numerator $\sigma_a\sigma_b$ in Eq. (\ref{ab2}) is a shorthand 
for the sum over all separations into two scatters of the source 
of their three jets. In this spirit, one could write more precisely
\begin{equation}
  \label{ab2-split1}
  \sigma_{\gamma+3\mathrm{jets};2} =
  \frac{\sigma_{\gamma+1\mathrm{jet}}\sigma_{2\mathrm{jets}}
    +\sigma_{\gamma+2\mathrm{jets}}\sigma_{1\mathrm{jet}}}{\mbox{\seffcdf}}\,.
\end{equation}
Note that the contribution containing $\sigma_{\gamma+0\mathrm{jets}}$,
which should in principle also appear here, was found to be negligible~\cite{Abe:1997xk}.

The data were corrected for the trigger efficiency, but not for any other detector 
effects, but we argue below (see Sec.~\ref{CorrectionFactor}) that, since we will only need
ratios of closely-related cross sections, they should largely cancel.

\label{CDF}

\section{Parton level correction}
In order to understand the correction at the parton level from \seffcdf{} to \seff{}, 
we write down the expression for the cross section $\sigma_{ab;2}$ according to 
Eq.~(\ref{ab2-split1}) in a simple eikonal model.  We assume that the
dependence of the parton distributions on the two-dimensional 
impact parameter, $\vec{b}$, and longitudinal momentum fraction, $x$, factorize.  
We make extensive use of the overlap function $A(\vec{b})$,
normalized such that
\begin{equation}
  \int\mathrm{d}^2b\;A(\vec{b})=1.
\end{equation}
To set the scene, we mention that the cross section for exactly $n$
scatters of a type $b$ is given~by
\begin{equation}
  \sigma_{b;n} = \int\mathrm{d}^2b\, \frac1{n!}\Bigl(\sigma_b\,A(\vec{b})\Bigr)^n
  \exp\biggl\{-\sigma_b\,A(\vec{b})\biggr\}.
\end{equation}
That is, the cross section is obtained by integrating over all values of
impact parameter the probability of the scatters: each has a probability
$\sigma_bA(\vec{b})$; there are $n$ of them and they are independent, giving the power of
$n$; and they are all of the same type, giving the $n!$ factor.
Finally, the exponential gives the probability that there are no other
scatters (of type $b$).  It is straightforward to check that the
inclusive cross section is given by
\begin{equation}
  \sigma_{b} = \sum_n n\,\sigma_{b;n}.
\end{equation}
For a total of $n$ scatters: exactly one of type $a$ and $n\!-\!1$ of
type $b$, with $a$ and $b$ assumed distinguishable, we obtain in the
same way
\begin{equation}
  \sigma_{ab;n} = \int\mathrm{d}^2b \Bigl(\sigma_a\,A(\vec{b})\Bigr)
  \,\frac1{(n-1)!}\Bigl(\sigma_b\,A(\vec{b})\Bigr)^{n-1}
  \exp\biggl\{-(\sigma_a+\sigma_b)A(\vec{b})\biggr\}.
\end{equation}
In all that follows, we assume that one of the two scattering types, let
us say $a$, has a small cross section, so we can drop $\sigma_a$ from
the exponent.  We therefore have, for the exactly two-scatter cross
section
\begin{equation}
  \sigma_{ab;2} = \sigma_a\,\sigma_b\int\mathrm{d}^2b
  \Bigl(A(\vec{b})\Bigr)^2 \exp\biggl\{-\sigma_b\,A(\vec{b})\biggr\},
\end{equation}
and hence
\begin{equation}
  \label{CDFEFF}
  \mbox{\seffcdf} = \frac1{\displaystyle{\int}\mathrm{d}^2b
  \Bigl(A(\vec{b})\Bigr)^2 \exp\biggl\{-\sigma_b\,A(\vec{b})\biggr\}}\,.
\end{equation}
If, instead, we define \seff{} through the inclusive two-scatter
cross section
\begin{equation}
  \sigma_{ab} = \sum_n (n-1)\sigma_{ab;n} =
  \sigma_a\,\sigma_b\int\mathrm{d}^2b \Bigl(A(\vec{b})\Bigr)^2,
\end{equation}
we obtain
\begin{equation}
  \label{EFF}
  \mbox{\seff} = \frac1{\displaystyle{\int}\mathrm{d}^2b \Bigl(A(\vec{b})\Bigr)^2}\,,
\end{equation}
independent of the cross sections for the individual processes selected.

It is clear that if the cross section for process $b$ is small, the two
definitions, Eqs.~(\ref{CDFEFF}) and (\ref{EFF}), become identical.
Keeping the first correction to this, we obtain
\begin{equation}
  \label{approx}
  \mbox{\seffcdf} \approx \mbox{\seff}+\R\sigma_b,
\end{equation}
where
\begin{equation}
  \R \equiv \frac{\displaystyle{\int}\mathrm{d}^2b \Bigl(A(\vec{b})\Bigr)^3}
  {\left[\displaystyle{\int}\mathrm{d}^2b \Bigl(A(\vec{b})\Bigr)^2\right]^2}
\end{equation}
is a function only of the \emph{shape}, but not the \emph{size}, of the
overlap function.  For example, for a Gaussian matter distribution we
have $\R=1.33$, for a matter distribution based on the electromagnetic
form factor, as used in Refs.\cite{Butterworth:1996zw, Bahr:2008dy}, we
have $\R=1.46$, for an exponential matter distribution, we have $\R=1.78$
and for a `black disk' we have $\R=1.26$ (we mention these numbers for
illustration, but they are not needed for our extraction).

Treleani's idea is to obtain $\R\sigma_b$ term from the rate of 3-scatter
events,
\begin{equation}
  \sigma_{ab;3} = \int\mathrm{d}^2b \Bigl(\sigma_a\,A(\vec{b})\Bigr)
  \,\frac1{2!}\Bigl(\sigma_b\,A(\vec{b})\Bigr)^2
  \exp\biggl\{-\sigma_b\,A(\vec{b})\biggr\}\,.
\end{equation}
With the same accuracy as we have just used to obtain
Eq.~(\ref{approx}), we can replace the exponential by unity and obtain
\begin{equation}
  \sigma_{ab;3} \approx \frac1{2!}\sigma_a\sigma_b^2\int\mathrm{d}^2b
  \Bigl(A(\vec{b})\Bigr)^3
  \approx \frac1{2!}\sigma_{ab;2}\,\frac{\R\sigma_b}{\mbox{\seff}}\,.
  \label{eq:sigma3_parton}
\end{equation}
By insertion of $\R$ from (\ref{eq:sigma3_parton}) into (\ref{approx}), we 
completely reproduce the correction suggested in \cite{Treleani:2007gi}:
\begin{equation}
  \mbox{\seffcdf} \approx \mbox{\seff}
  \left(1+2\frac{\sigma_{ab;3}}{\sigma_{ab;2}}\right).
\end{equation}

\label{PartonLevel}

\section{Jet level correction}

We turn now to a calculation of the same quantities, with the same
accuracy, but at the jet level using the event definition of CDF.
By jet level, we mean that we take account of the subsequent evolution
of the parton-level event into a multi-parton system and thence into the
hadrons that are reconstructed as jets in the detector. These steps
smear the results considerably, due to the emission of extra jets into
the event, the recoil of the primary jets from them, the smearing of jet
momenta by hadronization effects, the loss of hadronic energy out of the
jet, the merging of nearby jets, etc. For jet physics with a threshold
of 5~GeV, all of these effects are large. Not only are the jets smeared,
but it becomes impossible to match them up and know which jets came from
which scatters.

Returning to
Eq.~(\ref{ab2-split1}), i.e.~including the fact that the observed three
jets can come from two scatters in one of two ways, we obtain
\begin{eqnarray}
  \sigma_{\gamma+3\mathrm{jets};2} &=& \int\mathrm{d}^2b
  \biggl[\Bigl(\sigma_{\gamma+1\mathrm{jet}}\,A(\vec{b})\Bigr)
  \Bigl(\sigma_{2\mathrm{jets}}\,A(\vec{b})\Bigr) \nonumber \\
  &+&\Bigl(\sigma_{\gamma+2\mathrm{jets}}\,A(\vec{b})\Bigr)
  \Bigl(\sigma_{1\mathrm{jet}}\,A(\vec{b})\Bigr)\biggr]
  \exp\biggl\{-\sigma_{\mathrm{jets}}\,A(\vec{b})\biggr\}\,.
\end{eqnarray}
Note that the cross section in the exponent is a third type of
scattering: in both separations of events, we veto all scatters with
\emph{any} jets above 5~GeV.  Therefore $\sigma_{\mathrm{jets}}$ is the
cross section for a scatter to produce \emph{at least one\/} jet with
$E_T>5$~GeV.  With the same accuracy as the argumentation made above, we
therefore have
\begin{equation}
\label{eq:approx:hadron}
  \mbox{\seffcdf} \approx \mbox{\seff}+\R\sigma_{\mathrm{jets}}.
\end{equation}
On the other hand, the three-scatter contribution to this cross section
is given by
\begin{equation}
  \sigma_{\gamma+3\mathrm{jets};3} = \frac1{2!}\int\mathrm{d}^2b
  \Bigl(\sigma_{\gamma+1\mathrm{jet}}\,A(\vec{b})\Bigr)
  \Bigl(\sigma_{1\mathrm{jet}}\,A(\vec{b})\Bigr)
  \Bigl(\sigma_{1\mathrm{jet}}\,A(\vec{b})\Bigr)
  \exp\biggl\{-\sigma_{\mathrm{jets}}\,A(\vec{b})\biggr\}\,.
\end{equation}
That is, we continue to neglect the $\gamma+0\,\mathrm{jets}$ term, so
three jets from three scatters can only come about from each of the
extra scatters contributing exactly one jet.  Approximating this in the
same way as above and putting everything together we obtain
\begin{eqnarray}
  \label{neweqn}
  \mbox{\seffcdf} &\approx& \mbox{\seff}
  \left(1+2\frac{\sigma_{\gamma+3\mathrm{jets};3}}{\sigma_{\gamma+3\mathrm{jets};2}}
  \times\frac{\sigma_{\gamma+1\mathrm{jet}}\sigma_{2\mathrm{jets}}
    +\sigma_{\gamma+2\mathrm{jets}}\sigma_{1\mathrm{jet}}}
  {\sigma_{\gamma+1\mathrm{jet}}\sigma_{1\mathrm{jet}}}
  \times\frac{\sigma_{\mathrm{jets}}}{\sigma_{1\mathrm{jet}}}
  \right) \\
   &=& \mbox{\seff}
  \left(1+2\frac{\sigma_{\gamma+3\mathrm{jets};3}}{\sigma_{\gamma+3\mathrm{jets};2}}
  \left(\frac{\sigma_{2\mathrm{jets}}}{\sigma_{1\mathrm{jet}}}
    +\frac{\sigma_{\gamma+2\mathrm{jets}}}{\sigma_{\gamma+1\mathrm{jet}}}
      \right)
  \frac{\sigma_{\mathrm{jets}}}{\sigma_{1\mathrm{jet}}}
  \right) \\
   &=& \mbox{\seff}
  \left(1+2\frac{\sigma_{\gamma+3\mathrm{jets};3}}{\sigma_{\gamma+3\mathrm{jets};2}}
  \F
  \right),
\end{eqnarray}
where in the last equation we define the correction factor
\begin{equation}
\label{eq:f}
  \F = \left(\frac{\sigma_{2\mathrm{jets}}}{\sigma_{1\mathrm{jet}}}
    +\frac{\sigma_{\gamma+2\mathrm{jets}}}{\sigma_{\gamma+1\mathrm{jet}}}
      \right)
  \frac{\sigma_{\mathrm{jets}}}{\sigma_{1\mathrm{jet}}}\,.
\end{equation}
Note that at the parton level, in which scatters can be
identified and counted perfectly and jets are replaced by partons, the correction
factor $\F$  is equal to unity since each
of the two new ratios that have appeared in Eq.~(\ref{neweqn}) is equal
to one, since in this case $\sigma_{1\mathrm{jet}}$,
$\sigma_{2\mathrm{jets}}$ and $\sigma_{\mathrm{jets}}$ are all equal and
$\sigma_{\gamma+2\mathrm{jets}}$ is negligible.  In general, however, the
first new factor can be expected to be smaller than unity and the second
one larger than unity so it is not clear the direction of the overall
effect. 

We do not believe it is possible to extract these ratios from numbers in
the CDF paper alone. Nevertheless, since they are
ratios of closely-related cross sections, we may hope that they are
considerably better predicted than the cross sections themselves.  In
the next section we use Monte Carlo data to extract the correction factor 
and make an estimate of its uncertainty.

\label{HadronLevel}

\section{Estimate of the jet correction factor $\F$}
Before proceeding to the numerical estimate, we comment on the accuracy
required. Since the ratio (\ref{eq:s3overs2}) has an uncertainty of 
$\sim$ $^{+30}_{-50}\%$, an uncertainty of the correction factor $\F$ 
that multiplies it of order $20\%$ or less would be ample, leaving the result for the 
effective cross section dominated by the experimental uncertainties.
It would be unreasonable to aim for a significantly higher accuracy than this, 
since $\alpha_S(5\, \mathrm{GeV}) \sim 0.2$ and this analysis relies on the leading 
order Monte Carlo generators.

In order to calculate the correction factor $\F$ and more reliably estimate its accuracy, 
we used three different Monte Carlo event generators \HerwigL{}~\cite{Gieseke:2011na, Bahr:2008pv}, 
\fHerwigL{}~\cite{Corcella:2000bw,Corcella:2002jc} and \PythiaL{}~\cite{Sjostrand:2006za} 
to produce events of both $\gamma$~+~jets and pure QCD  jets types. Each generated event 
was handed over to the Rivet package~\cite{Buckley:2010ar} to be analyzed. This ensured that 
the computation of observables is exactly the same for each generator. Since the CDF analysis 
was not available among the standard Rivet's analyses we have implemented its event selection 
criteria (see Section~\ref{CDF}) into the package. For jet finding, we used the default settings 
of \CDFJetClu{} jet algorithm as implemented in \textsf{FastJet~2.4.2}~\cite{Cacciari:2011ma,Cacciari:2005hq} 
with a cone radius $R=0.7$, which according to~\cite{Abe:1991ui} was used in the original CDF analysis.  

In the CDF analysis, jet cuts depend on the order of jet in the event with respect to 
its transverse energy. The fact that the leading jet could be above or below the $7$~GeV, 
while the two trailing jets have to be below this threshold, induces an additional correlation 
between the two scatters, in principle. The sum over the two divisions of the jet origin between 
the two scatters should be extended to include a sum over all assignments of the leading jet between 
the two scatters, and whether it is above or below $7$~GeV. However, the part of the cross section 
coming from events in which the leading jet comes from the QCD scattering and not from the photon
production was found to be tiny, about $1.5\%$. Moreover, it is about the same fraction in the numerator 
and the denominator for the appropriate ratio, so neglecting these events really has a negligible effect. 
Therefore, the highest E$_T$ jet in $\gamma$ + jets is required to be above $5$~GeV and all other 
jets to be between $5$~GeV and $7$~GeV.

Since we want to make predictions for the final state of a single scattering, we switch off all MPI effects, 
but leave all other generator options on and at their default values.
As we have emphasised, jets of such low transverse momentum are strongly
smeared and only weakly correlated with the partonic scattering that produced
them. Correspondingly, there is a significant probability that jets that enter
our acceptance may be initiated by partonic scatterings with transverse
momentum well below that acceptance. In order to estimate the amount of this
migration, which also gives an idea of the contamination of the jet sample by
very soft physics, we divide the calculated cross sections into two separate
parts, which we call Soft and Hard, according to the matrix element~$\hat{p_t}$.
For the jets ($\gamma$~+~jets, respectively) production cross sections the Hard part is defined with 
$\hat{p_t}>2$~GeV ($\hat{p_t}>10$~GeV) and the Soft\footnote{It is worth noting that default setting of \Pythia{} 
does not produce any events with $\hat{p_t}<1.0$~GeV.} with $0.5<\hat{p_t}<2$~GeV ($5.0<\hat{p_t}<10$~GeV).

The obtained cross sections, their ratios and values of the correction factor for each generator are presented 
in Table~\ref{tab:main_table}\footnote{The statistical errors are negligible, therefore we suppress them in the Table.}.
The first observation is that the $\gamma$ + jets sample for $E_{T\gamma}>16$~GeV is well behaved: there is 
almost no migration from hard processes below $10$~GeV. However, in the case of the QCD scattering cross sections 
we see a significant amount of migration from below $2$~GeV.
We conclude that these events are somewhat less likely to be well
modeled, which is also reflected by significantly different cross sections obtained from different generators. 
Nevertheless, the amount of migration is similar in each of these cross sections, 
therefore the mis-modeling cancels to some extent in their ratios 
and hence the correction factor is reasonably predicted.
This is particularly evident when comparing the correction factors obtained
from \Herwig{} and \Pythia{}. Despite significant differences in the  
cross sections, the values of the correction factors are very close.
The reason why this coefficient is different in the case of \fHerwig{} will be explained 
later in this section. 

\begin{table}[fh]
\centering
\begin{tabular}{|c|c|c|c|c|c|c|c|c|c|}
\hline
  & \multicolumn{3}{|c|}{\Herwig{}} & \multicolumn{3}{|c|}{\fHerwig{}} & \multicolumn{3}{|c|}{\Pythia{}} \\
\hline \hline
$\sigma$ [mb]    & ~Hard~  & ~Soft~ & ~Sum~              & ~Hard~ & ~Soft~ & ~Sum~             & ~Hard~  & ~Soft~ & ~Sum~ \\
\hline
$\sigma_{1jet}$  &  9.16 & 3.16 & 12.32                      & 5.33 & 6.61 & 11.94                     & 6.93  & 2.51 & 9.44 \\
$\sigma_{2jets}$ &  0.62 & 0.15 &  0.77                      & 0.54 & 0.70 &  1.24                     & 0.72  & 0.00 & 0.72 \\
$\sigma_{jets}$  & 13.87 & 3.70 & 17.57                      & 8.72 & 8.31 & 17.03                     & 10.54 & 2.52 & 13.06 \\
\hline

$\sigma$ [nb]    & \multicolumn{3}{|c|}{~~~~~~~~}     & \multicolumn{3}{|c|}{~~~~~~~~}   & \multicolumn{3}{|c|}{~~~~~~~~} \\
\hline
$\sigma_{\gamma+1jet}$  &  5.66 & 0.03 & 5.69               & 3.41 & 0.16 & 3.57                     & 4.47  & 0.08 & 4.55\\
$\sigma_{\gamma+2jets}$ &  1.46 & 0.01 & 1.47               & 1.02 & 0.04 & 1.06                     & 1.05  & 0.07 & 1.22\\
\hline
\hline
$\frac{\sigma_{2jets}}{\sigma_{1jet}}$               & \multicolumn{3}{|c|}{0.063} & \multicolumn{3}{|c|}{0.103} & \multicolumn{3}{|c|}{0.076} \\
$\frac{\sigma_{jets}}{\sigma_{1jet}}$                & \multicolumn{3}{|c|}{1.426} & \multicolumn{3}{|c|}{1.426} & \multicolumn{3}{|c|}{1.383} \\
$\frac{\sigma_{\gamma+2jets}}{\sigma_{\gamma+1jet}}$ & \multicolumn{3}{|c|}{0.258} & \multicolumn{3}{|c|}{0.300} & \multicolumn{3}{|c|}{0.246} \\
\hline
$\F$                                                   & \multicolumn{3}{|c|}{0.458} & \multicolumn{3}{|c|}{0.575} & \multicolumn{3}{|c|}{0.445}\\
\hline
$\F_{avg.}$                                             & \multicolumn{9}{|c|}{0.493} \\
\hline
\end{tabular}
\caption{The calculated cross sections, their ratios and values of the correction factor 
for the default settings of each generator \Herwig{}, \fHerwig{}, and \Pythia{}. }
\label{tab:main_table}
\end{table}

We have already begun to address the question of how the results depend on 
the Monte Carlo modeling using three different event generators with their 
default settings. In addition, we use the \Herwig{} generator with the default settings altered in order 
to determine how the details of the simulation affect the results. More details are given in the appendices. 
The most important effects for our studies
are the order of $\alpha_S$ (1-loop and 2-loops) used in the simulation, width of the Gaussian distribution of the 
intrinsic transverse momentum $k_T$ of the interacting partons (we studied three values $0$, $1$ and $2$~GeV) 
and the parton distribution functions.  In addition, since CDF did its studies at calorimeter level and we 
do it at particle level, we believe that there is value in using a different jet algorithm (\CDFJetClu{}, 
\PxCone{}\cite{Seymour:2006vv}\footnote{The minimal transverse energy threshold was changed from the default
value of $0.5$~GeV to $0.1$~GeV.} and \AntiK{}\cite{Cacciari:2008gp} as implemented in FastJet 2.4.2), 
to see how dependent we are on these fine details. In Table~\ref{tab:pdf_table} we present results 
obtained using three different PDF sets 
(\MRST{}~\cite{Martin:1998sq},  \CTEQ{}~\cite{Pumplin:2002vw} and  \MRSTLO{}~\cite{Sherstnev:2007nd}) 
and two different orders of $\alpha_S$. We can see that the 
impact of the PDF on the result is small. Similarly, the order of  $\alpha_S$ has little effect on 
the $\F{}$ factor.
\begin{table}[fh]
\centering
\begin{tabular}{|c|c|c|c||c|c|c|}
\hline
 PDF & { \MRST{}} & { \CTEQ{}} & { \MRSTLO{}} & $\alpha_S$ &{1-loop } & {2-loops }\\
\hline
$\F$ & {0.477} & {0.447} & {0.458} & $\F{}$  & {0.476} & {0.458}\\
\hline
\end{tabular}
\caption{The correction factors obtained using \Herwig{} 
with three different PDF sets \MRST{}, \CTEQ{} and \MRSTLO{} (default in \Herwig{}) and 
two different orders of $\alpha_S$, 1-loop and 2-loops (default in \Herwig{}).}
\label{tab:pdf_table}
\end{table}
The jet clustering algorithms, see Table \ref{tab:alg_table}, have slightly bigger influence on the outcome but still 
smaller then the uncertainty coming from the different Monte Carlo models, see Table~\ref{tab:main_table}. 
\begin{table}[fh]
\centering
\begin{tabular}{|c|c|c|c|c|c|c|}
\hline
Jet algorithm  & \multicolumn{2}{|c|}{\CDFJetClu{}} & \multicolumn{2}{|c|}{\PxCone{}} & \multicolumn{2}{|c|}{\AntiK{}} \\
\hline
$\F$  & \multicolumn{2}{|c|}{0.458} & \multicolumn{2}{|c|}{0.512} & \multicolumn{2}{|c|}{0.525} \\
\hline
\end{tabular} 
\caption{The correction factors obtained using \Herwig{}
with three different jet algorithms \PxCone{}, \AntiK{} and \CDFJetClu{} (used in the CDF analysis).}
\label{tab:alg_table}
\end{table}

By far, the dominant effect is due to the intrinsic $k_T$ modeling, therefore we have studied its influence in more detail 
using all three generators. 
The results from Table~\ref{tab:kt_table} explain why the results obtained using default settings of \Pythia{} and 
\Herwig{} are very similar (see Table~\ref{tab:main_table}). This is because the intrinsic momentum in both generators was tuned to experimental 
data and have by default similar value $k_T \sim 2$~GeV, while in \fHerwig{} it was not tuned to the data and by default is equal to $0$~GeV. 
Therefore, in this respect results from \Herwig{} and \Pythia{}  should be trusted more then from \fHerwig{}.
We also see (Table~\ref{tab:kt_table}) that all generators provide a similar value of $\F$ for the same $k_T$ value.

\begin{table}[fh]
\centering
\begin{tabular}{|c|c|c|c|}
\hline
$\F$            & $k_T = 0.0$~GeV 	& $k_T = 1.0$~GeV 	& $k_T = 2.0$~GeV \\
\hline \hline
\Herwig{} 	& 0.648    		& 0.582    		& 0.465 \\
\fHerwig{}		& 0.575     		& 0.619    		& 0.564 \\
\Pythia{} 		& 0.620     		& 0.590    		& 0.445 \\
\hline
\end{tabular}
 \caption{The correction factors obtained using three generators and 
three different values of intrinsic $k_T$. }
\label{tab:kt_table}
\end{table}
The more detailed results including the cross sections and their ratios are included in Appendix \ref{Appendix}.

As an estimate of the correction factor we take the average value of $\F{}$ obtained from the
three different event generators with their default settings (see Table~\ref{tab:main_table}).
The systematic error of the estimation is taken as a half of the difference between the maximum 
and minimum value of $\F$ caused by the effects studied in this section. Therefore, the final result is
\begin{equation}
f_{avg.}=0.49\pm0.10\ .
\end{equation}
Which indeed, as we anticipated from the estimation of $\alpha_s(\mbox{$5$~GeV})$, is around 
$20$\% of the correction factor.

As a final cross-check, we calculate the fraction of
$\gamma$~+~$3$jets events that come from a $\gamma$~+~$2$jets collision
plus a $1$jet collision for which the experimental value was quoted as $\approx75$\%\cite{Abe:1997xk}.
We obtain 80\% in \Herwig{}, 74\% in \fHerwig{} and 76\% in \Pythia{}. 
Taking into account the inherent uncertainties in jet physics at $5$~GeV, the fact that we are working at
particle level and they work at uncorrected detector level, and the
accuracy we are aiming for in the final correction factor, we consider
this to be very good agreement with the experimental number.

\label{CorrectionFactor}

\section{Result}
Using the result of the previous section and the numbers in CDF's paper,
we are ready to extract a value for \seff{}.  However, we first
make one final comment, concerning the systematic errors.  In CDF's
analysis, the correction for triple-scattering events makes one of the
biggest single contributions to the final systematic error, since they
subtract the roughly estimated number of triple-scattering events off the number
of double-scattering events. We revise CDF's method and define a new effective cross section,
\seffcdfnew{}, which is the value of \seff{} they would
have obtained by keeping triple-scattering events in their sample and thus we avoid this source of the systematic error.  
Their master formula is
\begin{equation}
  \mbox{\seffcdf} = \frac{N_{DI}}{N_{DP}}
  \left(\frac{A_{DP}}{A_{DI}}\right) R_c \, \sigma_{NSD},
  \label{eq:seffcdf:experiment}
\end{equation}
with
\begin{eqnarray}
N_{DI} &=& 1060 \pm 110 \pm 110,  \\
N_{DP+TP} &=& 8865 \pm 430 \pm 150, \\
R_{DP} &=& 0.83 \pm 0 _{-0.04}^{+0.08},\\
N_{DP} &=& N_{DP+TP} \times R_{DP} = 7360 \pm 360 _{-380}^{+720}, \\
A_{DP}/A_{DI} &=& 0.958\pm0\pm0, \\
R_c &=& 2.06\pm0.02^{+0.01}_{-0.13}, \\
\sigma_{NSD} &=& (50.9\pm0\pm1.5)\mbox{~mb}.
\end{eqnarray}
where we have consistently written the first error as statistical and
the second as systematic, even when they are assumed to be zero. 
$N_{DI}$ stands for the number of double hadron interactions (DI) identified using
the vertex detector, $N_{DP}$ is the number of pure exclusive double parton scattering (DP)
events, whose fraction within all measured multiple-scattering events ($N_{DP+TP}$) is $R_{DP}$.
Factors $A_{DP}$ and $A_{DI}$ characterize the acceptances of the appropriate kinematic selections,
except the requirements connected to vertex reconstruction. Coefficient $R_c$ represents the ratio between the number
of beam crossings in which the detector found only one vertex and the number of beam crossings in which the detector 
reconstructed exactly two vertices. All beam crossings are taken into consideration in which any kind of non-single-diffractive (NSD)
inelastic proton--antiproton collision was detected. The appropriate cross section $\sigma_{NSD}$ is also given above.

Our new effective cross section has the same definition as in (\ref{eq:seffcdf:experiment}) but with $N_{DP}$ replaced by
$N_{DP+TP}$,
\begin{equation}
  \bar\sigma_{EFF,CDF} = \frac{N_{DI}}{N_{DP+TP}}
  \left(\frac{A_{DP}}{A_{DI}}\right) R_c \, \sigma_{NSD}
  = (12.0\pm1.4^{+1.3}_{-1.5})\mbox{~mb}.
\end{equation}
Note that the fractional systematic errors are indeed significantly
smaller.

Finally, we are ready to calculate \seff{}.  In terms of
\seffcdfnew{} it is given by
\begin{equation}
  \mbox{\seffcdfnew} = \mbox{\seff}
  \left(1+\frac{\sigma_{\gamma+3\mathrm{jets};3}}{\sigma_{\gamma+3\mathrm{jets};2}
  +\sigma_{\gamma+3\mathrm{jets};3}}
  \left[2
  \F_{avg.}
  -1\right]
  \right).
\end{equation}
\noindent Given that correction factor $\F_{avg.}$ is 0.49 $\pm$ 0.10, the factor in square brackets turns out to 
be very close to zero indicating that the difference between the inclusive measurement and the true inclusive
cross section for double parton scattering is very small.
The uncertainty on the triple-scattering event fraction 0.17$^{+0.04}_{-0.08}$ can be neglected with respect to the 
uncertainty of the $2\F_{avg.}-1$ term. Our final result for the effective cross section is
\begin{equation}
 \mbox{\seff} = (12.0 \pm 1.4 ^{+1.3}_{-1.5})~\mbox{mb}.
\end{equation} 
\noindent We see that despite the additional uncertainty coming from the correction factor, 
the systematic uncertainty is smaller and more symmetrical than on the CDF's final value.

\label{Results}

\section{Conclusions}
The CDF measurement of double parton scattering~\cite{Abe:1997xk} used a non-standard 
definition of \seff{} that makes this important quantity process dependent. Therefore,
the value provided by the experiment
$
 \mbox{\seffcdf}  = (14.5\pm1.7^{+1.7}_{-2.3})~\mbox{mb}
$
is not suitable for comparisons with other measurements or as input for theoretical calculations
or Monte Carlo models. The non-standard definition used by CDF has been pointed out and corrected for the first time by
Treleani in his publication~\cite{Treleani:2007gi}. Based on the theoretically-pure (parton-level) situation, \cite{Treleani:2007gi}
estimated the inclusive (process independent)
$
  \mbox{\seff}=10.3~\mbox{mb}.
$
This result would be correct under the assumption that CDF were able to uniquely identify 
and count the number of scatters in an event, which is certainly not the case.
In this publication we have considered CDF's event definition in more detail to provide an improved correction leading to
\begin{equation}
 \mbox{\seff} = (12.0 \pm 1.4 ^{+1.3}_{-1.5})~\mbox{mb}.
\end{equation} 
It is worth noting that both statistical and systematic uncertainties have decreased, since the additional uncertainty 
of our correction factor is much smaller than the avoided uncertainty stemming from the triple scattering removal done originally by CDF.

The obtained value of \seff{} serves as a constraint on the Monte Carlo models since the recent tunes of MPI
models to the LHC data predict its value to be between $25-42$~mb~\cite{Gieseke:2011xy}. 
This inconsistency between theory and experiment can also be seen from Eq.~\ref{eq:approx:hadron}, if one
uses the $\sigma_{\mathrm{jets}}$ around $15$~mb, see Table \ref{tab:main_table}, and $\R$ from Section~\ref{PartonLevel}.
This behavior suggests that the overlap function used in the MC models is oversimplified and should be improved, for example, 
by including $x$-dependence~\cite{Strikman:2011cx,Frankfurt:2010ea,Corke:2011yy}. This value can also help to understand 
the recent results from the LHCb experiment~\cite{Aaij:2012dz}(see page 23, Fig 10).  
The experimental results for \seff{} extracted from the production of $J/\psi$ mesons together with an associated open
charm hadron and from double open charm hadron production are different by a factor of between two and three.

\acknowledgments
We are grateful to the Cloud Computing for Science and Economy project (CC1) 
at IFJ PAN (POIG 02.03.03-00-033/09-04) in Cracow whose resources were used 
to carry out all the numerical calculations for this project. Thanks also to 
Mariusz Witek for his help with CC1 and Michael Albrow for his input on CDF analysis. 
This work was funded in part (AS and MHS) by the Lancaster-Manchester-Sheffield
Consortium for Fundamental Physics under STFC grant ST/J000418/1, in
part (MHS) by an IPPP Associateship, in part (MM) by RVO68407700
and in part by the MCnet FP6 Marie Curie Research Training Network
MRTN-CT-2006-035606 and the MCnetITN FP7 Marie Curie Initial Training
Network PITN-GA-2012-315877.
\clearpage
\appendix
\section{Detailed Results}
\label{Appendix}
\subsection{Intrinsic $k_T$ dependence}
\begin{table}[fh]
\centering
\begin{tabular}{|c|c|c|c|c|c|c|}
\hline
  & \multicolumn{2}{|c|}{\Herwig{}} & \multicolumn{2}{|c|}{\fHerwig{}} & \multicolumn{2}{|c|}{\Pythia{}} \\
\hline \hline
$\sigma$ [mb]    & ~~Hard~~  & ~~Soft~~               & ~~Hard~~ & ~~Soft~~              & ~~Hard~~  & ~~Soft~~ \\
\hline
$\sigma_{1jet}$  & 5.13  & 1.40                       & 5.33 & 6.61                      & 4.79  & 0.06 \\
$\sigma_{2jets}$ & 0.65  & 0.26                       & 0.54 & 0.70                      & 0.66  & 0.00 \\
$\sigma_{jets}$  & 8.67  & 2.05                       & 8.72 & 8.31                      & 8.04  & 0.06 \\
\hline

$\sigma$ [nb]    & \multicolumn{2}{|c|}{~~~~~~~~}     & \multicolumn{2}{|c|}{~~~~~~~~}   & \multicolumn{2}{|c|}{~~~~~~~~} \\
\hline
$\sigma_{\gamma+1jet}$  & 5.38 & 0.06                 & 3.41 & 0.16                       & 4.46  & 0.08 \\
$\sigma_{\gamma+2jets}$ & 1.39 & 0.01                 & 1.02 & 0.04                       & 0.92  & 0.15 \\
\hline
\hline
$\frac{\sigma_{2jets}}{\sigma_{1jet}}$               & \multicolumn{2}{|c|}{0.139} & \multicolumn{2}{|c|}{0.103} & \multicolumn{2}{|c|}{0.136} \\
$\frac{\sigma_{jets}}{\sigma_{1jet}}$                & \multicolumn{2}{|c|}{1.641} & \multicolumn{2}{|c|}{1.426} & \multicolumn{2}{|c|}{1.668} \\
$\frac{\sigma_{\gamma+2jets}}{\sigma_{\gamma+1jet}}$ & \multicolumn{2}{|c|}{0.256} & \multicolumn{2}{|c|}{0.300} & \multicolumn{2}{|c|}{0.236} \\
\hline
$\F$                                                    & \multicolumn{2}{|c|}{0.648} & \multicolumn{2}{|c|}{0.575} & \multicolumn{2}{|c|}{0.620} \\
\hline
\end{tabular}
\caption{The calculated cross sections, their ratios and the final correction factors for the intrinsic $k_T$ RMS = 0.0 GeV
for three MC generators. \CDFJetClu{} jet algorithm was used.}
\label{Tab:Intkt_zero}
\end{table} 
\begin{table}[fh]
\centering
\begin{tabular}{|c|c|c|c|c|c|c|}
\hline
  & \multicolumn{2}{|c|}{\Herwig{}} & \multicolumn{2}{|c|}{\fHerwig{}} & \multicolumn{2}{|c|}{\Pythia{}} \\
\hline \hline
$\sigma$ [mb]    & ~~Hard~~  & ~~Soft~~               & ~~Hard~~ & ~~Soft~~              & ~~Hard~~  & ~~Soft~~ \\
\hline
$\sigma_{1jet}$  & 5.72  & 0.86                       & 5.60 & 6.96                      & 4.78  & 0.10 \\
$\sigma_{2jets}$ & 0.64  & 0.08                       & 0.54 & 0.72                      & 0.75  & 0.00 \\
$\sigma_{jets}$  & 9.37  & 1.11                       & 9.01 & 8.88                      & 8.01  & 0.10 \\
\hline

$\sigma$ [nb]    & \multicolumn{2}{|c|}{~~~~~~~~}     & \multicolumn{2}{|c|}{~~~~~~~~}   & \multicolumn{2}{|c|}{~~~~~~~~} \\
\hline
$\sigma_{\gamma+1jet}$  & 5.50 & 0.09                 & 3.46 & 0.11                       & 4.48  & 0.08 \\
$\sigma_{\gamma+2jets}$ & 1.43 & 0.01                 & 1.13 & 0.06                       & 0.92  & 0.07 \\
\hline
\hline
$\frac{\sigma_{2jets}}{\sigma_{1jet}}$               & \multicolumn{2}{|c|}{0.109} & \multicolumn{2}{|c|}{0.100} & \multicolumn{2}{|c|}{0.136} \\
$\frac{\sigma_{jets}}{\sigma_{1jet}}$                & \multicolumn{2}{|c|}{1.591} & \multicolumn{2}{|c|}{1.424} & \multicolumn{2}{|c|}{1.668} \\
$\frac{\sigma_{\gamma+2jets}}{\sigma_{\gamma+1jet}}$ & \multicolumn{2}{|c|}{0.257} & \multicolumn{2}{|c|}{0.335} & \multicolumn{2}{|c|}{0.218} \\
\hline
$\F$                                                    & \multicolumn{2}{|c|}{0.582} & \multicolumn{2}{|c|}{0.619} & \multicolumn{2}{|c|}{0.590} \\
\hline
\end{tabular}
\caption{The calculated cross sections, their ratios and the final correction factors for the intrinsic $k_T$ RMS = 1.0 GeV
for three MC generators. \CDFJetClu{} jet algorithm was used.}
\label{Tab:Intkt_one}
\end{table} 
\clearpage
\begin{table}[fh]
\centering
\begin{tabular}{|c|c|c|c|c|c|c|}
\hline
  & \multicolumn{2}{|c|}{\Herwig{}} & \multicolumn{2}{|c|}{\fHerwig{}} & \multicolumn{2}{|c|}{\Pythia{}} \\
\hline \hline
$\sigma$ [mb]    & ~~Hard~~  & ~~Soft~~               & ~~Hard~~ & ~~Soft~~              & ~~Hard~~  & ~~Soft~~ \\
\hline
$\sigma_{1jet}$  & 9.73   & 3.79                       & 6.23  & 10.74                   & 6.93   & 2.51 \\
$\sigma_{2jets}$ & 0.65   & 0.13                       & 0.72  & 0.81                    & 0.72   & 0.00 \\
$\sigma_{jets}$  & 14.73  & 4.38                       & 10.87 & 12.99                   & 10.54  & 2.52 \\
\hline

$\sigma$ [nb]    & \multicolumn{2}{|c|}{~~~~~~~~}     & \multicolumn{2}{|c|}{~~~~~~~~}   & \multicolumn{2}{|c|}{~~~~~~~~} \\
\hline
$\sigma_{\gamma+1jet}$  & 5.51 & 0.05                 & 3.43 & 0.17                       & 4.47  & 0.08 \\
$\sigma_{\gamma+2jets}$ & 1.48 & 0.03                 & 1.08 & 0.05                       & 1.05  & 0.07 \\
\hline
\hline
$\frac{\sigma_{2jets}}{\sigma_{1jet}}$               & \multicolumn{2}{|c|}{0.058} & \multicolumn{2}{|c|}{0.090} & \multicolumn{2}{|c|}{0.076} \\
$\frac{\sigma_{jets}}{\sigma_{1jet}}$                & \multicolumn{2}{|c|}{1.414} & \multicolumn{2}{|c|}{1.404} & \multicolumn{2}{|c|}{1.383} \\
$\frac{\sigma_{\gamma+2jets}}{\sigma_{\gamma+1jet}}$ & \multicolumn{2}{|c|}{0.271} & \multicolumn{2}{|c|}{0.312} & \multicolumn{2}{|c|}{0.246} \\
\hline
$\F$                                                    & \multicolumn{2}{|c|}{0.465} & \multicolumn{2}{|c|}{0.564} & \multicolumn{2}{|c|}{0.445} \\
\hline
\end{tabular}
\caption{The calculated cross sections, their ratios and the final correction factors for the intrinsic $k_T$ RMS = 2.0 GeV
for three MC generators. \CDFJetClu{} jet algorithm was used.}
\label{Tab:Intkt_two}
\end{table} 
\subsection{Jet algorithm dependence}
\begin{table}[fh]
\centering
\begin{tabular}{|c|c|c|c|c|c|c|}
\hline
  & \multicolumn{2}{|c|}{\CDFJetClu{}} & \multicolumn{2}{|c|}{\PxCone{}} & \multicolumn{2}{|c|}{\AntiK{}} \\
\hline \hline
$\sigma$ [mb]    & ~~Hard~~  & ~~Soft~~               & ~~Hard~~ & ~~Soft~~              & ~~Hard~~  & ~~Soft~~ \\
\hline
$\sigma_{1jet}$  & 9.16 & 3.16                        & 10.90 & 3.91                      & 8.89   & 3.21 \\
$\sigma_{2jets}$ & 0.62 & 0.15                        & 0.88  & 0.24                      & 0.70   & 0.24 \\
$\sigma_{jets}$  & 13.87 & 3.70                       & 16.02 & 4.56                      & 13.38  & 3.98 \\
\hline

$\sigma$ [nb]    & \multicolumn{2}{|c|}{~~~~~~~~}     & \multicolumn{2}{|c|}{~~~~~~~~}   & \multicolumn{2}{|c|}{~~~~~~~~} \\
\hline
$\sigma_{\gamma+1jet}$  & 5.66 & 0.03                 & 5.28 & 0.04                      & 2.89  & 0.01 \\
$\sigma_{\gamma+2jets}$ & 1.46 & 0.01                 & 1.53 & 0.02                      & 0.83  & 0.00 \\
\hline
\hline
$\frac{\sigma_{2jets}}{\sigma_{1jet}}$               & \multicolumn{2}{|c|}{0.063} & \multicolumn{2}{|c|}{0.076} & \multicolumn{2}{|c|}{0.078} \\
$\frac{\sigma_{jets}}{\sigma_{1jet}}$                & \multicolumn{2}{|c|}{1.426} & \multicolumn{2}{|c|}{1.390} & \multicolumn{2}{|c|}{1.434} \\
$\frac{\sigma_{\gamma+2jets}}{\sigma_{\gamma+1jet}}$ & \multicolumn{2}{|c|}{0.258} & \multicolumn{2}{|c|}{0.292} & \multicolumn{2}{|c|}{0.288} \\
\hline
$\F$                                                    & \multicolumn{2}{|c|}{0.458} & \multicolumn{2}{|c|}{0.512} & \multicolumn{2}{|c|}{0.525} \\
\hline
\end{tabular}
\caption{The calculated cross sections, their ratios and the final correction factors in dependence on the jet clustering algorithm
using \Herwig{} generator.}
\label{Tab:Hpp_alg_dep}
\end{table} 
\clearpage
\begin{table}[fh]
\centering
\begin{tabular}{|c|c|c|c|c|c|c|}
\hline
  & \multicolumn{2}{|c|}{\CDFJetClu{}} & \multicolumn{2}{|c|}{\PxCone{}} & \multicolumn{2}{|c|}{\AntiK{}} \\
\hline \hline
$\sigma$ [mb]    & ~~Hard~~  & ~~Soft~~               & ~~Hard~~ & ~~Soft~~              & ~~Hard~~  & ~~Soft~~ \\
\hline
$\sigma_{1jet}$  & 5.33 & 6.61                        & 6.18 & 8.30                      & 5.73 & 9.58 \\
$\sigma_{2jets}$ & 0.54 & 0.70                        & 0.78 & 1.25                      & 0.71 & 1.96 \\
$\sigma_{jets}$  & 8.72 & 8.31                        & 9.85 & 10.72                     & 9.27 & 13.84 \\
\hline

$\sigma$ [nb]    & \multicolumn{2}{|c|}{~~~~~~~~}     & \multicolumn{2}{|c|}{~~~~~~~~}   & \multicolumn{2}{|c|}{~~~~~~~~} \\
\hline
$\sigma_{\gamma+1jet}$  & 3.41 & 0.16                 & 3.07 & 0.11                      & 1.64  & 0.02 \\
$\sigma_{\gamma+2jets}$ & 1.02 & 0.05                 & 1.05 & 0.06                      & 0.45  & 0.03 \\
\hline
\hline
$\frac{\sigma_{2jets}}{\sigma_{1jet}}$               & \multicolumn{2}{|c|}{0.103} & \multicolumn{2}{|c|}{0.140} & \multicolumn{2}{|c|}{0.175} \\
$\frac{\sigma_{jets}}{\sigma_{1jet}}$                & \multicolumn{2}{|c|}{1.426} & \multicolumn{2}{|c|}{1.421} & \multicolumn{2}{|c|}{1.510} \\
$\frac{\sigma_{\gamma+2jets}}{\sigma_{\gamma+1jet}}$ & \multicolumn{2}{|c|}{0.300} & \multicolumn{2}{|c|}{0.348} & \multicolumn{2}{|c|}{0.291} \\
\hline
$\F$                                                    & \multicolumn{2}{|c|}{0.575} & \multicolumn{2}{|c|}{0.693} & \multicolumn{2}{|c|}{0.704} \\
\hline
\end{tabular}
\caption{The calculated cross sections, their ratios and the final correction factors in dependence on the jet clustering algorithm
using \fHerwig{} generator.}
\label{Tab:Herwig_alg_dep}
\end{table} 
\begin{table}[fh]
\centering
\begin{tabular}{|c|c|c|c|c|c|c|}
\hline
  & \multicolumn{2}{|c|}{\CDFJetClu{}} & \multicolumn{2}{|c|}{\PxCone{}} & \multicolumn{2}{|c|}{\AntiK{}} \\
\hline \hline
$\sigma$ [mb]    & ~~Hard~~  & ~~Soft~~               & ~~Hard~~ & ~~Soft~~              & ~~Hard~~  & ~~Soft~~ \\
\hline
$\sigma_{1jet}$  & 9.62 & 2.88                        & 11.41 & 3.98                     & 8.84  & 2.50 \\
$\sigma_{2jets}$ & 1.17 & 0.00                        & 1.51  & 0.00                     & 1.16  & 0.00 \\
$\sigma_{jets}$  & 15.04 & 2.92                       & 17.20 & 4.02                     & 13.81 & 2.55 \\
\hline

$\sigma$ [nb]    & \multicolumn{2}{|c|}{~~~~~~~~}     & \multicolumn{2}{|c|}{~~~~~~~~}   & \multicolumn{2}{|c|}{~~~~~~~~} \\
\hline
$\sigma_{\gamma+1jet}$  & 4.47 & 0.08                 & 4.21 & 0.07                      & 1.90  & 0.02 \\
$\sigma_{\gamma+2jets}$ & 1.05 & 0.07                 & 1.04 & 0.07                      & 0.45  & 0.07 \\
\hline
\hline
$\frac{\sigma_{2jets}}{\sigma_{1jet}}$               & \multicolumn{2}{|c|}{0.076} & \multicolumn{2}{|c|}{0.085} & \multicolumn{2}{|c|}{0.075} \\
$\frac{\sigma_{jets}}{\sigma_{1jet}}$                & \multicolumn{2}{|c|}{1.383} & \multicolumn{2}{|c|}{1.338} & \multicolumn{2}{|c|}{1.378} \\
$\frac{\sigma_{\gamma+2jets}}{\sigma_{\gamma+1jet}}$ & \multicolumn{2}{|c|}{0.246} & \multicolumn{2}{|c|}{0.258} & \multicolumn{2}{|c|}{0.270} \\
\hline
$\F$                                                    & \multicolumn{2}{|c|}{0.445} & \multicolumn{2}{|c|}{0.459} & \multicolumn{2}{|c|}{0.475} \\
\hline
\end{tabular}
\caption{The calculated cross sections, their ratios and the final correction factors in dependence on the jet clustering algorithm
using \Pythia{} generator.}
\label{Tab:Pythia_alg_dep}
\end{table} 
\clearpage
\subsection{PDF dependence}
\begin{table}[fh]
\centering
\begin{tabular}{|c|c|c|c|c|c|c|}
\hline
 & \multicolumn{2}{|c|}{\MRST} & \multicolumn{2}{|c|}{\CTEQ} & \multicolumn{2}{|c|}{\MRSTLO} \\
\hline \hline
$\sigma$ [mb]    & ~~Hard~~  & ~~Soft~~               & ~~Hard~~ & ~~Soft~~              & ~~Hard~~  & ~~Soft~~ \\
\hline
$\sigma_{1jet}$  & 5.54  & 1.71                       & 6.39 & 1.73                      & 9.16   & 3.16 \\
$\sigma_{2jets}$ & 0.36  & 0.07                       & 0.41 & 0.07                      & 0.62   & 0.15 \\
$\sigma_{jets}$  & 8.46  & 1.98                       & 9.59 & 1.98                      & 13.87  & 3.70 \\
\hline

$\sigma$ [nb]    & \multicolumn{2}{|c|}{~~~~~~~~}     & \multicolumn{2}{|c|}{~~~~~~~~}   & \multicolumn{2}{|c|}{~~~~~~~~} \\
\hline
$\sigma_{\gamma+1jet}$  & 3.77 & 0.06                 & 4.21 & 0.04                      & 5.66  & 0.03 \\
$\sigma_{\gamma+2jets}$ & 1.00 & 0.04                 & 1.06 & 0.03                      & 1.46  & 0.01 \\
\hline
\hline
$\frac{\sigma_{2jets}}{\sigma_{1jet}}$               & \multicolumn{2}{|c|}{0.060} & \multicolumn{2}{|c|}{0.059} & \multicolumn{2}{|c|}{0.063} \\
$\frac{\sigma_{jets}}{\sigma_{1jet}}$                & \multicolumn{2}{|c|}{1.440} & \multicolumn{2}{|c|}{1.423} & \multicolumn{2}{|c|}{1.426} \\
$\frac{\sigma_{\gamma+2jets}}{\sigma_{\gamma+1jet}}$ & \multicolumn{2}{|c|}{0.271} & \multicolumn{2}{|c|}{0.255} & \multicolumn{2}{|c|}{0.258} \\
\hline
$\F$                                                    & \multicolumn{2}{|c|}{0.477} & \multicolumn{2}{|c|}{0.447} & \multicolumn{2}{|c|}{0.458} \\
\hline
\end{tabular}
\caption{The calculated cross sections, their ratios and the final correction factors in dependence on the parton distribution function
used in \Herwig{} generator.}
\label{Tab:Hpp_PDF_dep}
\end{table} 
\begin{table}[fh]
\centering
\begin{tabular}{|c|c|c|c|c|}
\hline
 &  \multicolumn{2}{|c|}{\CTEQL} & \multicolumn{2}{|c|}{\MRSTLO} \\
\hline \hline
$\sigma$ [mb]    & ~~Hard~~  & ~~Soft~~       & ~~Hard~~ & ~~Soft~~               \\
\hline
$\sigma_{1jet}$  & 9.62 & 2.88                       & 9.62  & 2.88                       \\
$\sigma_{2jets}$ & 1.17 & 0.00                       & 1.17  & 0.00                       \\
$\sigma_{jets}$  & 15.04 & 2.92                      & 15.04 & 2.92                       \\
\hline

$\sigma$ [nb]    & \multicolumn{2}{|c|}{~~~~~~~~}    & \multicolumn{2}{|c|}{~~~~~~~~}    \\
\hline
$\sigma_{\gamma+1jet}$  & 4.47 & 0.08                & 6.32 & 0.09                       \\
$\sigma_{\gamma+2jets}$ & 1.05 & 0.07                & 1.29 & 0.19                       \\
\hline
\hline
$\frac{\sigma_{2jets}}{\sigma_{1jet}}$               & \multicolumn{2}{|c|}{0.076} & \multicolumn{2}{|c|}{0.094} \\
$\frac{\sigma_{jets}}{\sigma_{1jet}}$                & \multicolumn{2}{|c|}{1.383} & \multicolumn{2}{|c|}{1.436} \\
$\frac{\sigma_{\gamma+2jets}}{\sigma_{\gamma+1jet}}$ & \multicolumn{2}{|c|}{0.246} & \multicolumn{2}{|c|}{0.232} \\
\hline
$\F$                                                    & \multicolumn{2}{|c|}{0.445} & \multicolumn{2}{|c|}{0.468} \\
\hline
\end{tabular}
\caption{The calculated cross sections, their ratios and the final correction factors in dependence on the parton distribution function
used in \Pythia{} generator}
\label{Tab:Pythia_PDF_dep}
\end{table} 
\clearpage
\subsection{Order of $\alpha_S$ dependence}
\begin{table}[fh]
\centering
\begin{tabular}{|c|c|c|c|c|}
\hline
 & \multicolumn{2}{|c|}{1-loop $\alpha_S$} & \multicolumn{2}{|c|}{2-loops $\alpha_S$}  \\
\hline \hline
$\sigma$ [mb]    & ~~Hard~~  & ~~Soft~~               & ~~Hard~~ & ~~Soft~~ \\
\hline
$\sigma_{1jet}$  & 13.99  & 5.23                       & 9.16  & 3.16    \\
$\sigma_{2jets}$ & 0.95   & 0.24                       & 0.62  & 0.15    \\
$\sigma_{jets}$  & 20.96  & 6.19                       & 13.87 & 3.70    \\
\hline

$\sigma$ [nb]    & \multicolumn{2}{|c|}{~~~~~~~~}     & \multicolumn{2}{|c|}{~~~~~~~~}   \\
\hline
$\sigma_{\gamma+1jet}$  & 6.53 & 0.05                 & 5.66 & 0.03               \\
$\sigma_{\gamma+2jets}$ & 1.77 & 0.04                 & 1.46 & 0.01               \\
\hline
\hline
$\frac{\sigma_{2jets}}{\sigma_{1jet}}$               & \multicolumn{2}{|c|}{0.062} & \multicolumn{2}{|c|}{0.063} \\
$\frac{\sigma_{jets}}{\sigma_{1jet}}$                & \multicolumn{2}{|c|}{1.412} & \multicolumn{2}{|c|}{1.426} \\
$\frac{\sigma_{\gamma+2jets}}{\sigma_{\gamma+1jet}}$ & \multicolumn{2}{|c|}{0.275} & \multicolumn{2}{|c|}{0.258} \\
\hline
$\F$                                                    & \multicolumn{2}{|c|}{0.476} & \multicolumn{2}{|c|}{0.458} \\
\hline
\end{tabular}
\caption{The calculated cross sections, their ratios and the final correction factors in dependence on the order of $\alpha_S$
used in \Herwig{} generator.}
\label{Tab:Hpp_alphas_dep}
\end{table}


% \begin{thebibliography}{99}

\bibliographystyle{JHEP}	% (uses file "plain.bst")
\bibliography{references}

% Please avoid comments such as "For a review'', "For some examples",
% "and references therein" or move them in the text. In general,
% please leave only references in the bibliography and move all
% accessory text in footnotes.

% Also, please have only one work for each \bibitem.

% \end{thebibliography}
\end{document}